\documentclass[twocolumn, twocolappendix]{aastex631}

\usepackage{mathrsfs}
\usepackage{amsmath}
\numberwithin{equation}{section}
\usepackage{graphicx}
\graphicspath{{images/}}
\usepackage{times}
\usepackage[varg]{txfonts}
\usepackage{hyperref}

\def\ngw{{N}}

\def\blu{{\mathcal{B}^L_U}}
\def\olu{{\mathcal{O}^L_U}}
\def\b{{\mathcal{B}}}
\def\bt{{\mathcal{B}_t}}
\def\ot{{\mathcal{O}_t}}

\def\rlu{{\mathcal{R}^L_U}}

\def\HL{{\mathcal{H}_L}}
\def\HU{{\mathcal{H}_U}}
\def\FPPpair{{\mathrm{\alpha}_\mathrm{pair}}}
\def\FPPcat{{\mathrm{\alpha}_\mathrm{cat}}}

\def\dt{{\Delta t}}

\begin{document}

\title{Lensing, not luck! Detection prospects of strongly lensed gravitational waves}

\author{Ankur Barsode$^a$}
\email{$^a$ankur.barsode@icts.res.in}
\affiliation{International Centre for Theoretical Sciences, Tata Institute of Fundamental Research, Bangalore 560089, India}

\author{Koustav N. Maity$^{b}$}
\email{$^c$koustav.narayan@icts.res.in}
\affiliation{International Centre for Theoretical Sciences, Tata Institute of Fundamental Research, Bangalore 560089, India}

\author{Parameswaran Ajith$^{c}$}
\email{$^b$ajith@icts.res.in}
\affiliation{International Centre for Theoretical Sciences, Tata Institute of Fundamental Research, Bangalore 560089, India}

\begin{abstract}
A small fraction of gravitational-wave (GW) signals detected by ground-based observatories will be strongly lensed by intervening galaxies or clusters. This may produce multiple copies of the signals (i.e., lensed images) arriving at different times at the detector. These, if observed, could offer new probes of astrophysics and cosmology. However, identification of lensed image pairs among a large number of unrelated GW events is challenging. Though the number of lensed events increases with improved detector sensitivity, the false alarms increase quadratically faster. While this ``lensing or luck'' problem would appear to be insurmountable, we show that the expected increase in measurement precision of source parameters will efficiently weed out false alarms. Based on current astrophysical models and anticipated sensitivities, we predict that the first confident detection could occur in the fifth observing run of LIGO, Virgo, and KAGRA. We expect computational costs to be a major hurdle in achieving such a detection, and show that the Posterior Overlap 2.0 method may offer a near-optimal solution to this challenge.
\end{abstract}


\section{Introduction}
\label{sec:introduction}

LIGO~\citep{aasi2015advanced} and Virgo~\citep{acernese2014advanced} have detected over two hundred gravitational-wave (GW) signals from merging binaries of black holes and neutron stars~\citep{abbott2019gwtc, abbott2021gwtc, abbott2023gwtc, abbott2024gwtc, abac2025gwtc, venumadhav2020new, zackay2021detecting, olsen2022new, mehta2025new, wadekar2023new, nitz20191, nitz20202, nitz20213, nitz20234, koloniari2025new}. In the upcoming observing runs, the number of GW detections will reach thousands~\citep{abbott2020prospects}, while the next-generation (XG) ground-based detectors are expected to observe millions of mergers~\citep{maggiore2020science, abac2025science, Reitze:2019iox}. A small fraction ($\sim 0.1\%$) of these GWs will be strongly lensed by intervening galaxies and clusters, producing multiple copies that will arrive at the detector with different time delays and magnifications~\citep{ng2018precise, xu2022please, wierda2021beyond, mukherjee2021impact, barsode2025fast}. Although several searches were performed on past data, no confirmed detections have been reported so far \citep{hannuksela2019search, dai2020search, mcisaac2020search, LIGOScientific:2021izm, abbott2023search, janquart2023follow}.

Observation of strongly lensed GWs promises new avenues for cosmography \citep{Liao:2017ioi,Wei:2017emo,jana2024strong}, probes of dark matter \citep{jana2025probing, barsode2024constraints}, and galaxy properties \citep{seo2024inferring, xu2022please}, and will enable new tests of general relativity \citep{Goyal:2020bkm, Goyal:2023uvm, Collett:2016dey, Fan:2016swi}. They may also be used to improve the localization of the GW source~\citep{hannuksela2020localizing, wempe2024detection, uronen2025finding} and to provide early warning to electromagnetic telescopes~\citep{Magare:2023hgs}. To enable this science, we must have robust and efficient methods to identify strongly lensed image pairs in GW data.

Under the geometric optics approximation, the waveforms of strongly lensed GWs are similar to those of unlensed GWs except for an overall magnification, a Morse phase shift of integer multiples of $\pi/2$, and a time delay \citep{dai2017waveforms, ezquiaga2021phase}. Therefore, most such strongly lensed signals can be \emph{detected} using the standard GW searches using unlensed waveform templates.

Despite this simplicity, it is challenging to \emph{identify} pairs of lensed GWs among the multitude of possible pairs in the detected signals, owing to the modest signal-to-noise ratio (S/N) of GW detections and the rarity of lensing events. Thus, lensing searches typically compute the Bayesian likelihood ratio between the ``lensed vs. unlensed'' hypotheses from every pair of detected signals (see \cite{barsode2025fast} for a review). However, noise fluctuations and accidental consistency between unlensed events can cause false alarms~\citep{ccalicskan2023lensing}. Thus, one puts a high enough threshold on the Bayesian likelihood ratio corresponding to an acceptably low false positive probability (FPP).

The issue of false alarms will worsen with the number of GW detections $N$: While the expected number of lensed image pairs (true positives) will grow $\propto N$, the number of unlensed pairs among all detections (false positives) will grow $\propto N^2$. This is the problem of ``lensing or luck''~\citep{ccalicskan2023lensing}. If, in the hope of encountering more lensed signals, one observes for a longer duration, it becomes \emph{harder} to distinguish between truly lensed events and (un)lucky unlensed pairs showing accidental consistency. \cite{ccalicskan2023lensing} argued that, in the upcoming observing runs of present-generation detectors, current search methods may not be enough to identify a lensed event before false alarms begin to dominate.

Presence of additional information, in the form of type-II lensed images~\citep{dai2017waveforms, ezquiaga2021phase, vijaykumar2023detection, janquart2021identification, wang2021identifying} or electromagnetic counterparts~\citep{smith2023discovering, smith2025multi, birrer2025challenges, ryczanowski2025follow, pastor2025fast}, could improve the situation~\citep{keitel2025false}. However, these effects are expected to be rare. One could also restrict the lensing search to a subspace of the signals' parameter space where lensed events are predominant but false alarms are less likely; e.g., signal pairs which are loud or have short time delays \citep{wierda2021beyond, ccalicskan2023lensing, chakraborty2025false}. \cite{hannuksela2025strong} have shown that time delay priors are particularly helpful in turning the growth of false alarms to $\propto \ngw$. However, their results hold only in the limit of large observing duration, or, one risks losing a considerable fraction of truly lensed signals.

Better detector sensitivity will increase not only the number of GW detections, but also the precision in the measurement of signal parameters. In this paper, we show that the improved precision will indeed compensate for the growth in false alarms in upcoming observing runs. We forecast that the first 3$\sigma$-confident detection of strongly lensed GWs is likely to happen in the fifth observing run (O5) of the LIGO-Virgo-KAGRA (LVK) detectors \citep{aasi2015advanced, acernese2014advanced, somiya2012detector, aso2013interferometer, akutsu2021overview}. With the anticipated upgrades of current detectors and with the advent of XG detectors, lensing detection shall become routine. Much of the exciting science cases proposed in the literature may kick-start in the coming decade.

Part of the reason why these forecasts are challenging is because lensing searches are computationally costly. Robust frequentist detection claims require simulating ``deep'' backgrounds (low FPP), which also becomes quadratically more expensive as we target larger $\ngw$. The forecasts reported in this paper are made possible by a computationally efficient Bayesian search method that we presented in \cite{barsode2025fast}, combined with a correspondence between the expected distributions of the Bayesian likelihood ratio under the \emph{lensed} and \emph{unlensed} hypotheses.

Section~\ref{sec:GWSL_search} provides an overview of the methods used to search for strongly lensed GWs and describes the challenges associated with their identification. In Section~\ref{sec:forecasts}, we describe our main results containing the detection forecasts, before concluding in Section~\ref{sec:conclusion}. We defer technical details to the Appendices~\ref{sec:GWSL_BB_plot}, \ref{sec:sim_strong_lensing}, and \ref{sec:astro_sim}.

\section{Overview of GW strong lensing search}
\label{sec:GWSL_search}
Given data $d_1$ and $d_2$ containing two GW signals, we can compute the Bayesian odds between the \emph{lensed} hypothesis $\HL$ (that the two signals are lensed copies of the same) and the \emph{unlensed} hypothesis $\HU$ (that the two signals are unrelated) as:
\begin{eqnarray}
\label{eq:posterior_odds}
\olu = \dfrac{P(\HL \mid d_1, d_2)}{P(\HU \mid d_1, d_2)} = \Pi^L_U ~ \blu,
\end{eqnarray}
where $\Pi^L_U$ denotes the prior odds\footnote{In this study, we will be in the limit of $\ngw > 100$, of which only a small fraction $u < 0.01$ may be lensed. Thus, we will write $\ngw(\ngw-1)/2 = \ngw^2/2$, $u/(1+u) = u$, etc.}~\citep{hannuksela2025strong}
\begin{equation}
\label{eq:prior_odds}
\Pi^L_U = \dfrac{P(\HL)}{P(\HU)} = \dfrac{2u}{\ngw},
\end{equation}
while $\blu$ is the Bayes factor that we compute from data. This is the likelihood ratio of getting a pair of signals $d_1,d_2$ under lensed vs. unlensed hypotheses
\begin{equation}
\label{eq:Bayes_factor}
\blu = \dfrac{P(d_1, d_2 \mid \HL)}{P(d_1, d_2 \mid \HU)}.
\end{equation}
All of the Bayesian search methods (e.g., \cite{haris2018identifying, janquart2021fast, janquart2023return, liu2021identifying, lo2023bayesian,goyal2024rapid, barsode2025fast}), compute $\blu$ or its approximation from the data (see, \cite{barsode2025fast} for a derivation). A large value of $\olu$ would favor the lensing hypothesis.

In practice, computing $\blu$ from a large number of event pairs is computationally expensive. However, \cite{barsode2025fast} have demonstrated an efficient method to compute $\blu$ using a set of realistic approximations (called \emph{Posterior Overlap 2.0}, or, PO2.0). This amounts to computing the inner product of the posterior distributions of the source parameters estimated from the two events (i.e., $d_1$ and $d_2$), appropriately weighted by their prior distributions informed by the population properties of the sources and lenses.

The likelihood ratio $\blu$ is also the optimal lensing identification statistic in the frequentist sense --- something that maximizes the identification efficiency ($\eta$) for a given value of FPP ($\FPPpair$) from a pair of GW signals~\citep{neyman1933ix}. Using the distribution $P(\blu \mid \HU)$ obtained empirically from a simulation of unlensed (background) pairs, we can compute the FPP of any unlensed pair having $\blu$ higher than the threshold $\bt$:
\begin{equation}
\label{eq:FPPpair}
\FPPpair(\bt) = \int\limits_{\bt}^\infty d\blu ~ P(\blu \mid \HU).
\end{equation}

Similarly, the efficiency of lensing identification (the fraction of lensed events present in the data that are correctly identified by the search pipeline at a given threshold) can be computed from the distribution $P(\blu \mid \HL)$ obtained from a simulation of lensed event pairs
\begin{equation}
\label{eq:efficiency}
\eta(\bt) = \int\limits_{\bt}^\infty d\blu ~ P(\blu \mid \HL).
\end{equation}

Since we typically search in a catalog of $\ngw$ signals, a more relevant FPP is that of finding \textit{at least one} pair with $\blu$ higher than the threshold in a catalog composed entirely of unlensed signals. Hence $\FPPpair$ has to be multiplied by the number of ``trials'', i.e., the number of pairs ($\ngw^2/2$), resulting in the ``catalog'' FPP~\citep{hannuksela2019search, ccalicskan2023lensing}
\begin{equation}
\label{eq:FPPcat}
\FPPcat(\bt) = \dfrac{\ngw^2}{2}~\FPPpair(\bt).
\end{equation}
The interpretation of $\FPPcat$ as a probability is valid only in the limit of $\FPPpair \ngw^2 \ll 1$. However, it  can always be interpreted as the expected number of false positives. In the frequentist picture, one may claim a detection of strong lensing if $\FPPcat$ is sufficiently small, which is often expressed in terms of Gaussian standard deviations, such as $3\sigma ~(\FPPcat \simeq 3\times10^{-3}) $ or $5\sigma ~(\FPPcat \simeq 6\times10^{-7}$).

The expected number of lensed events that are detectable at a given threshold $\bt$ is
\begin{equation}
N_\ell (\bt) = u \, \ngw \, \eta(\bt),
\label{eq:Nlens_det}
\end{equation}
where $u$ is the lensing fraction. At the same threshold, the expected number of false alarms is $\FPPcat(\bt)$. Thus, we can define a \emph{catalog purity} as the fraction of truly lensed events in a catalog of lensing candidates.
\begin{eqnarray}
\label{eq:purity}
\mathrm{Pty}(\bt) = \dfrac{u \, \ngw \, \eta(\bt)}{u \, \ngw \, \eta(\bt) + \FPPcat(\bt)} = \left({1 + \frac{N}{2u} \frac{\FPPpair(\bt)}{\eta(\bt)}}\right)^{-1}.
\end{eqnarray}
Impure catalogs with known purity are generally more useful than pristinely pure but less efficient ones (see, e.g.,~\cite{jana2024strong}). Thus, instead of the usual frequentist significance bounds, catalog purity itself may be used as a criteria for setting detection thresholds.

In Equation~\eqref{eq:purity}, ${N}/{2u} \equiv 1/\Pi^L_U$ will continue to increase in future observing runs as the number of GW detections will grow much faster than the lensing fraction $u$.  An optimal analysis pipeline tries to minimise ${\FPPpair(\bt)}/{\eta(\bt)}$. Due to the computational cost, it is difficult to get realistic estimates of ${\FPPpair(\bt)}/{\eta(\bt)}$ for most of the Bayesian lensing identification pipelines. However, thanks to the computational efficiency of the PO2.0 pipeline, \cite{barsode2025fast} has demonstrated a 70\% efficient identification at a $\FPPcat \equiv 2.3\sigma$ at current GW detector sensitivities (i.e. the projected sensitivity of the fourth observing run (O4) of LVK~\citep{HLVK-psd-O3O4O5}). However, since the expected rate of lensed events is low ($ u N \sim 0.2$) at current sensitivities, we must patiently wait until upgraded detectors come online in order to detect lensed events.

As GW detector networks improve, their higher sensitivity implies a higher S/N for a given source, allowing far-away sources to become detectable. This has both favorable and unfavorable effects on lensing searches:

\paragraph{Favorable effects}
\begin{enumerate}
\item[$\uparrow$] We detect more signals, leading to an increase in the number of lensed signals in the data $(u\ngw)$. There is also a small increase in the lensing fraction $u$ due to the deeper observing horizon.
\item[$\uparrow$] Thanks to the higher S/N and longer in-band duration of the signals, posterior distributions of source parameters shrink in size, reducing the probability of accidental consistency between unlensed pairs, which reduces $\FPPpair$.
\item[$\uparrow$] With additional detectors joining the network, the sky localization improves independently of S/N, further reducing $\FPPpair$.
\end{enumerate}

\paragraph{Unfavorable effects}
\begin{enumerate}
\item[$\downarrow$] The prior odds of lensing $(2u/\ngw)$ decrease because the increase in $u$ is negligible compared to the increase in $\ngw$.
\item[$\downarrow$] Time delay and magnification priors of lensed events broaden and become more similar to those of unlensed pairs, increasing $\FPPpair$ \citep{more2022improved}.
\end{enumerate}

The decrease in prior odds is the dominant effect that degrades the lensing identification efficiency, while the shrinkage of posteriors due to higher S/N is the dominant effect that counters it. If our network is so sensitive that we can observe all the GW sources in the universe, $\ngw$ will not increase further. In this limit, lensing search efficiency will only increase with further improvements in sensitivity.

However, for the intermediate sensitivities in the near future, it is not trivial to gauge whether these conflicting effects will lead to an overall improvement or degradation in efficiency. Even in the era of XG detectors, where the above limit may be realized, it is unclear whether they themselves will be sensitive enough to identify lensing against formidable prior odds of $\sim 10^{-8}$ (see Appendix~\ref{sec:astro_sim}), or whether even higher sensitivity is required.

We could answer these questions by simulating unlensed and lensed GW signals in various observing scenarios, computing their $\blu$'s, and empirically obtaining the distributions $P(\blu \mid \HU)$ and $P(\blu \mid \HL)$. These can then be used to estimate the pairwise FPP and efficiency using Equations~\eqref{eq:FPPpair} and \eqref{eq:efficiency} respectively. The challenge here is that as $\ngw$ increases, the number of unlensed simulations and $\blu$ calculations required to probe low values of $\FPPcat$ grows quadratically, making estimation of $P(\blu \mid \HU)$ computationally prohibitive (see Figure~\ref{fig:CPU_traditional}). In addition, the cost of estimating $P(\blu \mid \HL)$ from simulations of lensed events, though independent of $\ngw$, is still quite high.

In this study, we overcome these challenges using two key facts. First, under controlled conditions, there exists a relationship between $P(\blu \mid \HU)$ and $P(\blu \mid \HL)$, and we use it to bypass running expensive unlensed simulations (see Appendix~\ref{sec:GWSL_BB_plot} for details). Second, we obtain $P(\blu \mid \HL)$ expected in future observing runs by modeling it making use of the expected scaling of $\blu$ with S/N (see Appendix~\ref{sec:sim_strong_lensing} and \ref{sec:astro_sim} for details, including a description of observing scenarios assumed). Combined with the time delay prior ratio, this allows us to forecast strong lensing search efficiency at a manageable computational cost.

\section{Efficiency of GW strong lensing search}
\label{sec:forecasts}

\subsection{Efficiency of lensing identification}

\begin{figure}[t]
\centering
\includegraphics[width=\columnwidth]{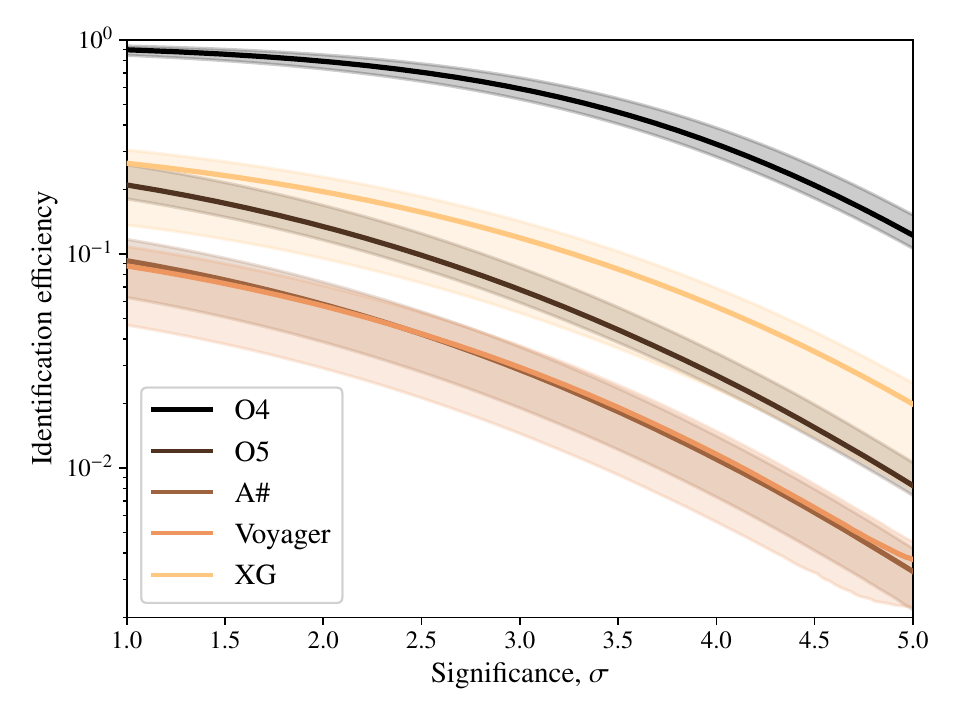}
\caption{The efficiency of identifying strongly lensed GWs plotted against the significance of detection for various observing scenarios. Solid lines show the expectation based on realistic assumptions of BBH merger rate and lensing fraction, while shaded regions show the range between more pessimistic and optimistic assumptions as defined in Appendix~\ref{sec:sim_strong_lensing}.}
\label{fig:future_efficiency_ROC}
\end{figure}

\begin{figure}[t]
\centering
\includegraphics[width=\columnwidth]{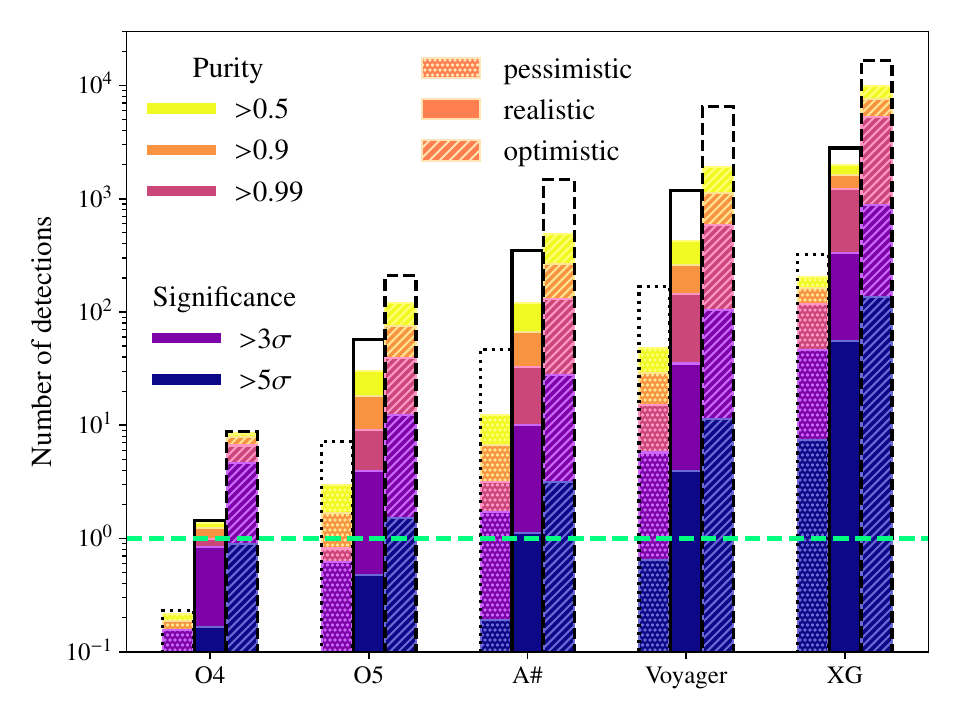}
\caption{The expected number of lensed GWs present in the data (unfilled black boxes) and the number of confident identifications (colored bars) in various observing scenarios. We show results for standard frequentist significance thresholds as well as those based on the purity of lensed catalogs. Results under pessimistic, realistic and optimistic assumptions of BBH merger rate and lensing fraction (as defined in Appendix~\ref{sec:sim_strong_lensing}) are differentiated using hatches and edge styles.}
\label{fig:future_efficiency_numbers}
\end{figure}

Figure~\ref{fig:future_efficiency_ROC} shows the lensing identification efficiency as a function of the catalog FPP. We find that the efficiency is highest at O4 sensitivity --- nearly 60\% (12\%) for $3\sigma~(5\sigma)$ detections. This is higher by an order of magnitude than O5. This could be because the lensing time delay distribution at O4 sensitivity is very different from that of unlensed pairs, as compared to other scenarios. This results in larger $\rlu$'s and $\blu$'s in spite of the on-average lower S/N (see Appendix~\ref{sec:sim_strong_lensing}). Secondly, the lower sensitivity means that the detection rate and number of false alarms (encoded by the $\ngw^2/2$ trials factor) are also lower. However, because of the low GW detection rate and lensing fraction, the \emph{number} of confident lensing identifications is not expected to be high in O4 (Figure~\ref{fig:future_efficiency_numbers}). A $3\sigma$ lensing identification in 3 years of observation at O4 sensitivity is realistically unlikely.

While optimistically there may be as many as four $3\sigma$ lensing detections in O4, the ongoing O4 run is scheduled to span only 2.5 years \citep{shoemaker2024observing}, so the intrinsic lensing rate will be lower than our estimate based on 3 years. Furthermore, while our simulation assumed a network of LIGO-Hanford, LIGO-Livingston, and Virgo detectors, the latter was not operational during the first leg (O4a), leading to degraded sky localization during that period. This significantly reduces the chances of a 3$\sigma$ detection since sky-localization consistency plays a major role in weeding out false alarms \citep{haris2018identifying, wong2021using, barsode2025fast}.

As the detector sensitivity improves beyond O4, initially, there will be a \emph{decrease} in identification efficiency. This is because the detector's observing horizon will reach the anticipated peak of the merger rate distribution~\citep{madau2014cosmic, dominik2013double}. The expected increase in the number of new detections, and hence the corresponding false alarms, overshadows the corresponding improvement in the parameter estimation. Time delays also become less useful at distinguishing lensed pairs as their distribution shifts towards higher values, becoming more and more similar to that of the unlensed pairs.

Once the detection horizon goes beyond the peak in merger rate distribution, we enter the era where improvement in sensitivity leads to only a slow increase in $\ngw$. The increase in trials factors slows down, and the ever increasing Bayes factors result in an increasing trend in lensing identification efficiency. We find that the efficiency will reach its minimum of around 3\% (0.3\%) for $3\sigma~(5\sigma)$ detections in the A\# detector, before rising again to about $10-20\% (1-3\%)$ for $3\sigma~(5\sigma)$ detections in the XG era, depending on the merger rate assumptions.

Multiplying the efficiency by the expected number of lensed events in the data [Equation~\eqref{eq:Nlens_det}], we plot the number of detections in three years in Figure~\ref{fig:future_efficiency_numbers}. Beyond O4, in spite of the initial decrease in detection efficiency, there will be an ample number of detections throughout various observing runs. Notably, our forecasts indicate that the first $3\sigma$ confident detection is most likely to occur in O5: About 4 events are expected in 3 years under realistic astrophysical model assumptions. Even pessimistically, the rate is $\sim$0.6, which is a $\sim$50\% chance of detection. Optimistically, there may be as many as twelve 3$\sigma$ detections, and perhaps even one 5$\sigma$ detection.

Beyond O5, lensing detection will become routine, with $10^{+18}_{-8}$, $35^{+68}_{-29}$, and $333^{+550}_{-287}$ $3\sigma$ detections in three years of observations in A\#, Voyager and XG detectors. On the other hand, 5$\sigma$ detections will be rarer, about $1^{+2}_{-1}$, $4^{+7}_{-3}$, and $56^{+81}_{-48}$, respectively. Figure~\ref{fig:future_efficiency_numbers} also shows the number of detections in catalogs of varying purity. Since these are more relaxed detection criteria, there are considerably more detections than traditional 3$\sigma$ or 5$\sigma$ thresholds. This is promising for downstream analyses, and we will return to this point in Section~\ref{sec:computational_cost}.

These forecasts are conservative since they do not consider improvements in sky localization independent of improving S/N, i.e. due to better GW networks. Secondly, these results are extrapolated from parameter estimation carried out using waveform templates modeling only aligned-spin effects and dominant modes of the gravitational radiation. In the future, parameter estimation will further improve due to inclusion of subdominant modes and spin precession effects, likely improving the lensing identification efficiencies.

\subsection{Computational efficiency}
\label{sec:computational_cost}
Confident identification of lensed events also poses a significant computational challenge --- both in the actual search for lensed pairs and in the background simulations used to establish the significance of the lensed candidates.

The current approach for lensing search adopted by the LVK collaboration is to use a multi-stage strategy \citep{wright2025lensingflow}, in which most of the less-significant event pairs are discarded using computationally inexpensive lensing search methods \citep{haris2018identifying, ezquiaga2023identifying}. A small fraction (optimistically $\sim 1\%$) of the event pairs that report the highest significance are then followed up by computationally expensive joint parameter estimation pipelines that calculate the Bayes factor given in Equation~\eqref{eq:Bayes_factor}. We optimistically assume that, for one pair of events, this analysis can be completed using $\sim 10$ CPU hrs of computing.

\begin{figure}[t]
\centering
\includegraphics[width=\columnwidth]{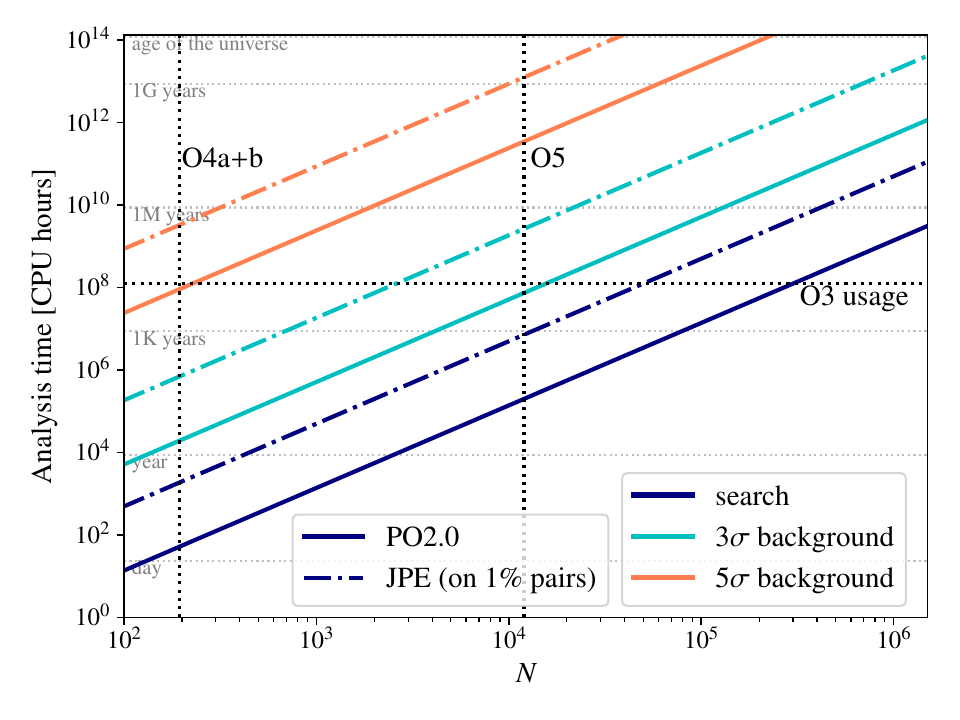}
\caption{The computational cost of various components of GW strong lensing search (i.e., that of computing $\blu$'s of all observed pairs), and background simulations needed to reach $3\sigma$ and $5\sigma$ significance (i.e., that of computing $\blu$'s of the required number of simulated unlensed pairs). Solid lines correspond to the analysis using PO2.0, while the dashed lines correspond to analysing 1\% of the event pairs using joint parameter estimation methods. For comparison, the total resource utilization of LVK's O3 run \citep{bagnasco2024ligo} is also shown.}
\label{fig:CPU_traditional}
\end{figure}

Now, to identify a lensed pair with a given value of $\FPPcat$, among $N$ GW events, we need to determine a $\blu$ threshold corresponding to a pairwise FPP of $\FPPpair = 2 \FPPcat/N^2$ [see Equation~\eqref{eq:FPPcat}]. Thus, in order to make a $3\sigma$ detection in O5 $(\FPPcat \simeq 3\times10^{-3}) $, with $\ngw \sim 10^4$, establishing confidence would require $\sim 10^5$ background simulations and $\sim 10^{10}$ $\blu$ calculations. The associated computational costs are shown in Figure~\ref{fig:CPU_traditional}. Using LVK's current strategy (where the computationally expensive joint parameter estimation is performed only on $\sim 1\%$ of the event pairs), a 3$\sigma$ detection in O5 will cost more than the resource utilization of \emph{all} the analyses performed in LVK's O3 run \citep{bagnasco2024ligo}. Achieving $5\sigma$ significance will be infeasible in O5 under current strategies.

Below, we now discuss some strategies for mitigating these costs:

\paragraph{1) Efficient Bayesian lensing identification methods:} ~The PO2.0 pipeline can compute the same Bayes factor, albeit with some approximations, in $\sim 10$ CPU seconds, saving over three orders of magnitude in computational cost. Its efficiency was demonstrated in \cite{barsode2025fast}. In Appendix~\ref{sec:GWSL_BB_plot} of this work, we show that it is also very close to the optimal Bayesian performance.

\paragraph{2) Frequentist significance estimates from lensed foregrounds:}~ Optimal Bayesian pipelines, using model assumptions that are close to Nature's ground truth, can use relatively-inexpensive simulations of \emph{lensed} events to estimate the frequentist significance of an event pair (see Appendices~\ref{sec:GWSL_BB_plot} and \ref{sec:OLU_mapping}). However, analysis pipelines must first verify the correctness of their implementations using the \emph{Bayes factor -- Bayes factor (B-B)} plots (see Appendix~\ref{sec:GWSL_BB_plot}), which in turn require background simulations. Moreover, real data may deviate from our model assumptions. We therefore do not recommend relying on this approach unless the uncertainties in astrophysical models are substantially reduced and the detector noise is well understood.

\paragraph{3) Catalog purity instead of frequentist significance:}~Traditional frequentist criteria require that lensed catalogs be almost perfectly pure, with chances of contamination as low as $\sim 10^{-3}$ for $3\sigma$. The first lensing identification will probably demand such a high degree of confidence. However, as we detect more and more lensed signals, and focus shifts towards doing further science using them, such stringent criteria may be wasteful. For example, \cite{jana2024strong} shows that even a 50\% pure catalog, containing equal number of lensed and unlensed pairs (but not knowing which pair is which), will enable excellent lensing cosmography.

Thus we could adopt catalog purity itself as a detection criterion. This leads to a considerable computational advantage: for $\mathcal{O}(1)$ purity and efficiency, Equation~\eqref{eq:purity} implies that one only needs to probe $\FPPpair$ accurately down to $\sim u/N$. The required number of background simulations then scales $\propto N$, reducing the computational cost by orders of magnitude in future observing runs.

Equation~\eqref{eq:purity} tells that estimation of catalog purity also requires simulations of lensed events, which rely on prior assumptions on the lens properties and population. However, unlike unlensed simulations whose costs grow $\propto \ngw^2$, the costs for lensed simulations are independent of $\ngw$, and are generally much lower. Thus, one can simply run multiple lensed simulations with different priors to assess their effects.

\section{Conclusion}
\label{sec:conclusion}

When searching for strongly lensed GWs, the number of false positives increases quadratically with the number of GW detections, as compared to the number of true lensed events which grow linearly. While this ``lensing or luck'' problem would appear to be insurmountable, we show that the expected increase in measurement precision of source parameters would allow us to outrun the false alarms. Based on current astrophysical models and anticipated sensitivities, we forecast that the first $3\sigma$ detection of lensing is likely to happen in O5, and it will be followed by a steady growth of high-purity catalogs. At XG sensivities, we can expect several hundreds of confident lensing detections per year, few tens exceeding $5\sigma$.

These forecasts assume \cite{madau2014cosmic} redshift distribution of binary black hole mergers fitted to \cite{abbott2023population} inferred population, using the population of galaxies as lenses. Lensing detection remains feasible under pessimistic assumptions and the rate may be $2-3$ times higher under optimistic ones. Improvements in sky localization due to better GW detector networks and measurement of subdominant modes of GW radiation may lead to even higher detection rates.

These forecasts are for pairwise searches using GW data alone. These may be further complemented by detections of individual type-II images or multi-messenger counterparts. There is also a possibility that some of the strongly lensed images may exhibit additional wave optics distortions due to stars and star clusters present in the lens galaxy~\citep{mishra2024exploring, chan2025detectability}. If detected, these distorted signals may in turn improve prospects of detecting strong lensing.

Computational cost of background simulations (which will grow quadratically with the number of GW events) is a major hurdle to making confident detections. The same hurdle lies in the way of making forecasts, though we circumvent it in this work by: 1) using the computationally efficient and near-optimal Bayesian lensing identification pipeline PO2.O~\citep{barsode2025fast}, 2) using the expected scaling of the Bayes factors with S/N to predict the Bayes factor distribution of lensed events expected in future observing runs, and 3) using the correspondence between Bayes factor distribution of unlensed events and lensed events.

Even with efficient Bayesian or machine-learning methods~\citep{goyal2021rapid, magare2024slick, li2025identification, campailla2025machine, offermans2024using}, finding thresholds corresponding to conventional $3\sigma$ or $5\sigma$ detections will be prohibitive in XG era due to the need of expensive background simulations. We propose to use the purity of the lensed catalog for determining identification thresholds. Though prior dependent, such thresholds are considerably less computationally expensive to estimate, and may prove valuable for downstream science using lensed GWs. A reanalysis of open data from LIGO-Virgo for lensing identification is currently underway.

The techniques presented here can also be applied to forecast the efficiency of identifying lensed binary neutron star mergers as well. These would likely be higher than for binary black holes due to the signals being longer and their parameters better constrained. Together with lensed binary black holes, these promise an exciting future for astrophysics and cosmology based on strongly lensed GWs.

\section*{Acknowledgments}
We thank Justin Janquart for a careful review of this manuscript. We acknowledge his and Srashti Goyal's role in motivating alternative detection thresholds, and the Infosys Excellence grant for enabling that meeting. We are grateful to Siva Athreya, Tejaswi Venumadhav, Uddeepta Deka, Anjali Bhatter, Soumav Kapoor, Otto Hannuksela, and Rico Lo for valuable discussions. We also thank members of the Astrophysical Relativity group at ICTS and the LVK Lensing group for constructive feedback on this work. Our research is supported by the Department of Atomic Energy, Government of India, under Project No. RTI4001. The numerical calculations reported in the paper were performed on the Alice computing cluster at ICTS-TIFR.

This work makes use of the \texttt{cogwheel} \citep{roulet2022removing, islam2022factorized}, \texttt{scipy} \citep{2020SciPy-NMeth}, \texttt{bilby} \citep{bilby_paper}, and \texttt{pyCBC} \citep{alex_nitz_2024_10473621} software packages.

\appendix

\section{Bypassing unlensed simulations}
\label{sec:GWSL_BB_plot}
Here we describe how to bypass the computationally expensive evaluation of the distribution Bayes factors under the unlensed hypothesis, $P(\blu | \HU)$. This is achieved by establishing a correspondence between the conditional distribution of the Bayes factor under competing hypotheses.

Given the competing hypotheses $\HL$ and $\HU$ (which include the signal, noise, and population assumptions), the Bayes factor $\blu$ defined in Equation~\eqref{eq:Bayes_factor} is a real valued function of the two data vectors $d_1$ and $d_2$. Thus, the expected distribution of $\blu$ is given by
\begin{equation}
P(\blu) = \int {d}d_1~{d}d_2~P(d_1, d_2)~\delta\left[\blu-\blu(d_1,d_2)\right],
\end{equation}
where $\delta$ denotes the Dirac delta function and $\blu(d_1,d_2)$ is given by Equation~\eqref{eq:Bayes_factor}. Then, the conditional distribution of the Bayes factor under $\HL$ is
\begin{equation}
\label{eq:likelihood_transform_HL}
P(\blu \mid \HL) = \int {d}d_1 \, {d}d_2~P(d_1, d_2 \mid \HL) \, \delta\left[\blu-\blu(d_1,d_2)\right],
\end{equation}
whereas the same under $\HU$ is
\begin{equation}
P(\blu \mid \HU)  =  \int {d}d_1 \, {d}d_2~P(d_1, d_2 \mid \HU) \, \delta\left[\blu-\blu(d_1,d_2)\right].
\end{equation}
Using Equation~\eqref{eq:Bayes_factor}, this can be rewritten as
\begin{equation}
\label{eq:likelihood_transform_HU}
P(\blu \mid \HU) = \int {d}d_1 \, {d}d_2~\frac{P(d_1, d_2 \mid \HL)}{\blu}\, \delta\left[\blu-\blu(d_1,d_2)\right].
\end{equation}
Comparing Equations~\eqref{eq:likelihood_transform_HL} and \eqref{eq:likelihood_transform_HU}, it is easy to see that
\begin{equation}
\label{eq:blu_distribution_relationship_inverted}
P(\blu \mid \HU) = \dfrac{P(\blu \mid \HL)}{\blu}.
\end{equation}

This result is significant: it allows us to obtain the distribution of $\blu$ under the unlensed hypothesis without performing costly background simulations. We instead run the relatively inexpensive lensed simulations, and construct $P(\blu \mid \HU)$ by reweighting those $\blu$ samples by $\blu$ itself. This may seem odd at first: what is connecting the results of two seemingly independent simulations? The answer lies in the fact that the $\blu$ function necessarily incorporates the population priors, as well as the signal and noise models of \emph{both the hypotheses}. Conversely, this also means that Equation~\eqref{eq:blu_distribution_relationship_inverted} holds only when the data distribution matches the assumptions made in the calculation of $\blu$.

In the analysis of real GW signals, waveform systematics, noise artifacts, and population model uncertainties may result in the data $d_1,d_2$ being distributed differently from the assumptions that enter $\blu$ calculation. Simulating an unlensed background then becomes necessary for a robust detection of strong lensing. However, for forecasts using simulated data as done in this paper, we can ensure that the model assumptions match between the hypotheses and the $\blu$ calculation.

\begin{figure}[t]
\centering
\includegraphics[trim={0 0 13.9cm 0},clip,width=\columnwidth]{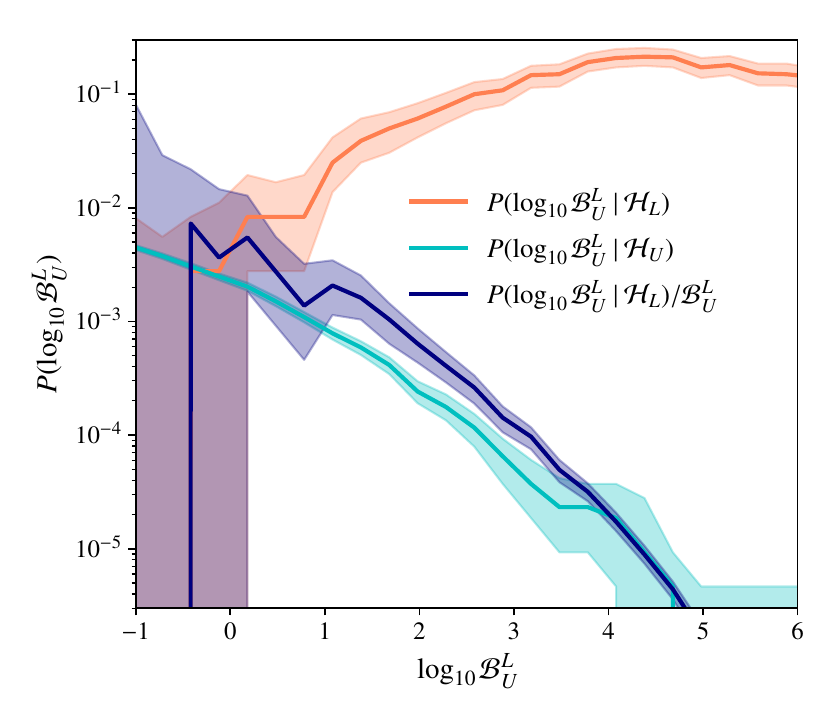}
\caption{The probability density of simulated strong lensing Bayes factors under the lensed and unlensed hypotheses, $P(\log_{10}\blu \mid \HL)$, $P(\log_{10}\blu \mid \HU)$, along with $P(\log_{10}\blu \mid \HL) / \blu$. Shaded regions show $90\%$ multinomial error bounds. These $\blu$'s, computed using \cite{barsode2025fast}'s PO2.0 method, verify Equation~\eqref{eq:blu_distribution_relationship_inverted} within an accuracy of a factor of 2.}
\label{fig:BB_plot}
\end{figure}

Figure~\ref{fig:BB_plot} provides a numerical validation of this relation using $\blu$'s computed from simulated lensed and unlensed events, using the PO2.0 pipeline. The small discrepancy that exists between $P(\log_{10}\blu \mid \HU)$  and $P(\log_{10}\blu \mid \HL) / \blu$ is likely due to the choice of settings in kernel density reconstructions used in the PO2.0 method. Slightly different choices shift the distributions slightly, though the discrepancy remains $\lesssim 2$, or $\lesssim 0.7$ in $\ln\blu$. Note that even joint parameter estimation techniques that directly compute $\blu$ from $d_1, d_2$ typically have a statistical error of $0.3$ in $\ln\blu$, coming from individual errors in one lensed and two unlensed evidence calculations. An implementation of PO2.0 with minimal reconstruction errors is currently in progress. We stress that the presence of this bias \emph{does not} contradict \cite{barsode2025fast}'s results --- the frequentist FPP (Equation~\ref{eq:FPPpair}) is left unchanged by an overall bias in $\blu$.

Under controlled conditions, Equation~\eqref{eq:blu_distribution_relationship_inverted} can be used to formulate a diagnostic test for Bayesian model selection pipelines. This involves performing simulations where the data is drawn from the competing hypotheses $\HL$ and $\HU$ and computing the distributions of the Bayes factor in each cases, i.e., $P(\blu | \HL)$ and $P(\blu | \HU)$. The ratio $P(\blu | \HL) / P(\blu | \HU)$  should be equal to $\blu$ itself. Any discrepancies would indicate incorrect formulation or coding bugs. This is similar in spirit to the probability-probability ($p-p$ plots) that are commonly employed to check the validity of Bayesian parameter estimation pipelines (see, e.g., \cite{dax2021real}).

\begin{figure}[t]
\centering
\includegraphics[width=\columnwidth]{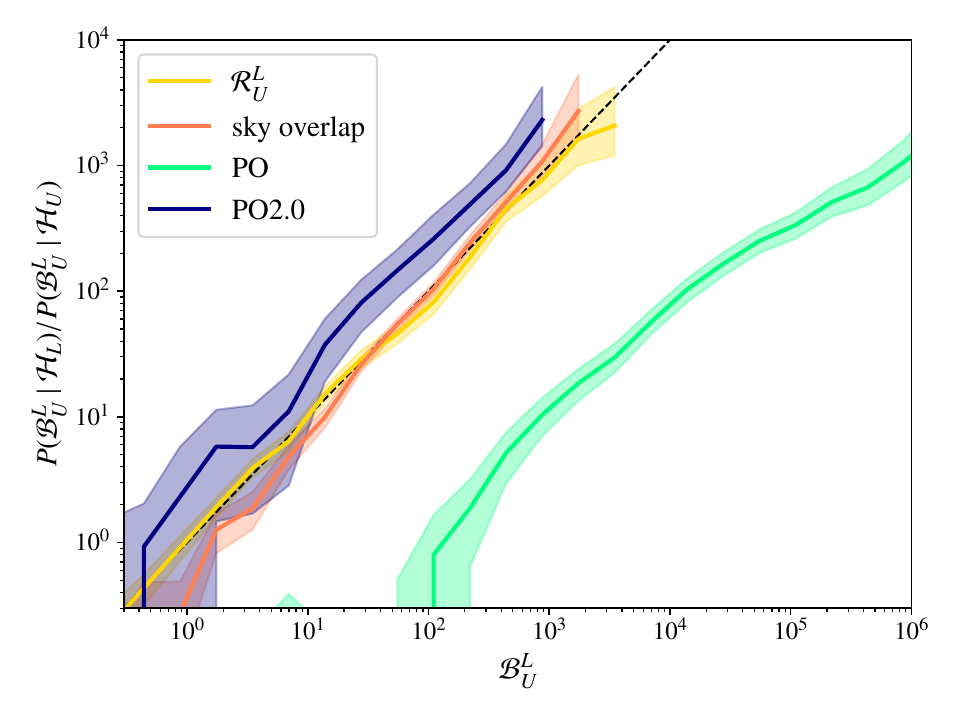}
\caption{The Bayes factor-Bayes factor ($B-B$) plot showing $P(\blu | \HL) / P(\blu | \HU)$ as a function of $\blu$ for different Bayesian model selection methods: 1) \cite{haris2018identifying}'s $\rlu$ incorporating consistency of time delays, 2) sky localization overlap, 3) \cite{haris2018identifying}'s PO method taking posterior overlap of selected BBH parameters assuming flat priors, and 4) \cite{barsode2025fast}'s PO2.0 method utilizing all the available information from dominant-mode aligned-spin posteriors with correct astrophysical priors.}
\label{fig:BB_plot_other_stats}
\end{figure}

Figure~\ref{fig:BB_plot_other_stats} shows the ``$B-B$ plots" of PO2.0 along with a few other Bayes factor formulations of lensing identification from literature. Firstly, we find that \cite{haris2018identifying}'s $\rlu$ statistic, which checks for the consistency of time delays with the lensed and unlensed hypotheses, has a perfectly diagonal $B-B$ plot. This is not really surprising since in this case the variable transformation from equation~\eqref{eq:likelihood_transform_HL} becomes one-to-one, which has very little numerical error.

The same is true for the sky overlap $\mathcal{S}^L_U$ (equation B1 of~\cite{barsode2025fast}), which is a specific case of \cite{haris2018identifying}'s overlap statistic for the angular location of the source in the sky). Note that the right normalization is important to give the right conclusions under a Bayesian interpretation ($4\pi$ as done in \cite{barsode2025fast} as opposed to $8\pi^3$ in \cite{goyal2021rapid}), although this would not affect the frequentist significance.

Finally, we find that the $B-B$ plot for \cite{haris2018identifying}'s full Posterior Overlap statistic is heavily biased, and the bias is not simply a constant. This Bayes factor is computed without population priors or selections effects (called the ``coherence ratio'' in joint parameter estimation pipelines such as \cite{janquart2021fast} and \cite{lo2023bayesian}), which leads to an overestimation by as much as two orders of magnitude \citep{cheung2023mitigating}. If these coherence ratios were interpreted on face value under a Bayesian detection philosophy, one would find too many false alarms.

Apart from the issues related to incorrect formulation or implementation errors discussed above, there is also the problem that real GW data may not satisfy our model assumptions, leading to biased results even from correct implementations. Thus, there is considerable risk involved in interpreting Bayes factors at face value, i.e., without an accompanying background. However, if a pipeline has demonstrated a correct $B-B$ plot, then, at least in controlled conditions, Equation~\ref{eq:blu_distribution_relationship_inverted} can be used to skip running background simulations. This is true for the PO2.0 pipeline to a good extent.

\section{Bypassing lensed simulations in future observing runs}
\label{sec:sim_strong_lensing}
To begin, we note that, because arrival times -- and therefore time delays -- are measured with negligible error, the strong lensing Bayes factor $\blu$ can be written as \citep{haris2018identifying, barsode2025fast}
\begin{equation}
\label{eq:BLU_split_to_RLU}
\blu = \b \times \rlu
\end{equation}
where $\b$ is the Bayes factor computed over BBH and lensing parameters excluding the arrival time and time delay, but conditioned on the observed time delay of the pair, and $\rlu$ is the ratio of time delay priors under the lensed and unlensed hypotheses evaluated at the pair's time delay~\citep{haris2018identifying}
\begin{equation}
\label{eq:rlu}
\rlu = \dfrac{P(\dt \mid \HL)}{P(\dt \mid \HU)}.
\end{equation}

\begin{figure}[t]
\centering
\includegraphics[width=\columnwidth]{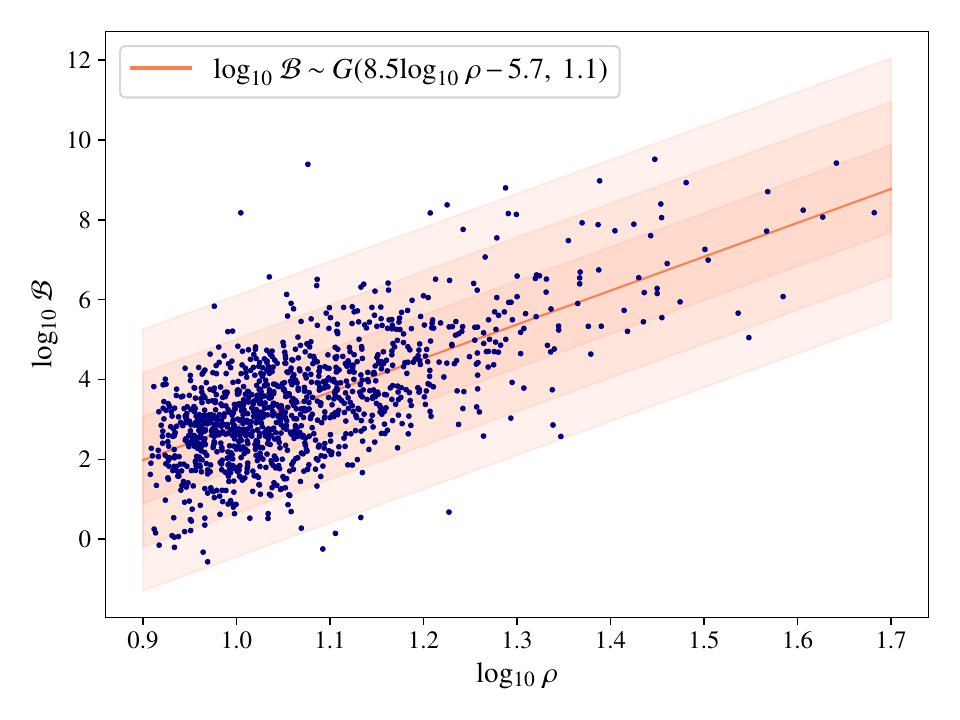}
\caption{Distribution of $\log_{10}\b$ ($=\blu / \rlu$) plotted against the effective S/N for simulated lensed events. This distribution is well fit by Gaussians with a fixed standard deviation and a mean that varies linearly with $\log_{10}\rho$. The middle six standard deviations of these Gaussians are shown using shaded regions.}
\label{fig:B0_vs_rho0}
\end{figure}

The $\rlu$'s can be computed with relatively inexpensive simulations, since there is no need to perform parameter estimation here. However, parameter estimation is needed for evaluating $\b$'s for a large number of lensed GW injections, which is a computationally expensive task. Instead of running simulations in each future observing scenario, we will use an expected scaling of the Bayes factor with the S/N of the lensed images to extrapolate results from a single simulation set at current sensitivity. Assuming that the posteriors can be approximated by multivariate Gaussians, the expected scaling is~\citep{gao2023identifying}
\begin{equation}
\label{eq:B0_vs_SNR0_scaling}
\b = b \left(\frac{\rho}{\rho_0} \right)^D, ~~~ \mathrm{with}~
\rho = \left({\dfrac{2\rho_1^2\rho_2^2}{\rho_1^2+\rho_2^2}}\right)^{1/2},
\end{equation}
where $\rho_1$ and $\rho_2$ are the S/N of the two individual events and $\rho_0$ is some characteristic S/N (say, 8). We empirically estimate the scaling exponent $D$ (which is expected to be close to the dimensionality of the posterior distributions that are used to compute $\b$) and the scaling constant $b$ from simulations assuming O4 sensitivity. Note that noise will cause some scatter on $b$ (see Figure~\ref{fig:B0_vs_rho0}). Thus, we model $\log_{10} b$ as belonging to a Gaussian distribution of mean $\mu$ and standard deviation $\sigma$. We then estimate the maximum likelihood values of $\mu, \sigma$ and $D$ assuming a Gaussian likelihood~\footnote{We find the best fit value of $D$ to be 8.5, which is quite close to the effective dimensionality of the parameter space that is used to compute $\b$, i.e., masses ($m_1, m_2$) and spin magnitudes ($a_1, a_2$) of the black holes, sky location ($\alpha, \delta$), polarization angle and coalescence phase (which are completely degenerate, thus form one effective dimension $\psi - \phi_0$), luminosity distance and inclination angle (highly correlated, and hence the effective dimensionality is between 1 and 2).}. This allows us to model the distribution of $\b$ as a function of the S/N. In Figure~\ref{fig:B0_vs_rho0}, the blue dots show the actual values of $\b$ and the orange shaded regions show the contours of the modeled distribution. This should hold, to a good approximation, in future observing runs also. To predict the $\b$ expected in future observing runs, we simulate a population of BBHs, identify the detectable events after computing their S/N, and draw random values of $\b$ from the modeled distribution.

\section{Astrophysical simulations of lensing in future observing scenarios}
\label{sec:astro_sim}

\begin{table*}[t]
\centering
\caption{Description of various current and upcoming GW detector networks that we consider in this study, adopted from Maity et al. (in prep). Locations of LIGO, Virgo, KAGRA, and ET are adopted from \cite{lalsuite}, while for CE they are taken from \cite{borhanian2021gwbench}. The number of observed signals in 3 years, $\ngw$, and lensing fraction, $u$, are a result of applying selection effects under each of these scenarios to astrophysical populations of BBH sources following \cite{abbott2023population}, \cite{madau2014cosmic}, lensed by galaxies distributed according to the SDSS catalog \citep{collett2015population}.}
\renewcommand{\arraystretch}{1.1}
\begin{tabular}{ccccc}
\\ \hline \hline
\begin{tabular}[c]{@{}c@{}}Scenario\end{tabular} &
Detector locations (sensitivity) & \begin{tabular}[c]{@{}c@{}}References to noise curves\end{tabular} & $\ngw$ & $u$ [\%] \\ \hline

O4 & \begin{tabular}[c]{@{}c@{}}LIGO Hanford (O4a), \\ LIGO Livingston (O4a), \\ Virgo Italy (O4a)\end{tabular} & \begin{tabular}[c]{@{}c@{}}\cite{HLVK-psd-O4a} \\ \cite{HLVK-psd-O4a} \\ \cite{HLVK-psd-O4a}\end{tabular} & $4.5\times10^2$ & 0.38 \\ \hline

O5 & \begin{tabular}[c]{@{}c@{}}LIGO Hanford (A+), \\ LIGO Livingston (A+), \\ Virgo Italy (AdV+), \\ KAGRA Japan (KAGRA+)\end{tabular} & \begin{tabular}[c]{@{}c@{}}\cite{HLVK-psd-O3O4O5} \\ \cite{HLVK-psd-O3O4O5} \\ \cite{HLVK-psd-O3O4O5} \\ \cite{HLVK-psd-O3O4O5}\end{tabular} & $9.6\times10^3$ & 0.49 \\ \hline

A\# & \begin{tabular}[c]{@{}c@{}}LIGO Hanford (A\#), \\ LIGO Livingston (A\#), \\ LIGO India (O3)\end{tabular} & \begin{tabular}[c]{@{}c@{}}\cite{HL-psd-asharp} \\ \cite{HL-psd-asharp} \\ \cite{HLVK-psd-O3O4O5}\end{tabular} & $6\times10^4$ & 0.46 \\ \hline

Voyager & \begin{tabular}[c]{@{}c@{}}LIGO Hanford (Voyager), \\ LIGO Livingston (Voyager), \\ LIGO India (A+)\end{tabular} & \begin{tabular}[c]{@{}c@{}}\cite{HL-psd-voyager} \\ \cite{HL-psd-voyager} \\ \cite{HLVK-psd-O3O4O5}\end{tabular} & $3\times10^5$ & 0.44 \\ \hline

XG & \begin{tabular}[c]{@{}c@{}}LIGO Hanford (Voyager), \\ LIGO Livingston (Voyager), \\ LIGO India (A+), \\ Einstein Telescope Italy (ET-D, 10 km xylophone), \\ Cosmic Explorer US (CE1[40 km], CE2[20 km])\end{tabular} & \begin{tabular}[c]{@{}c@{}}\cite{HL-psd-voyager} \\ \cite{HL-psd-voyager} \\ \cite{HLVK-psd-O3O4O5} \\ \cite{borhanian2021gwbench} \\ \cite{Reitze:2019iox}\end{tabular} & $1.5\times10^6$ & 0.53
\\ \hline \hline
\end{tabular}
\label{tab:scenario_definitions}
\end{table*}

To compute $\blu$'s of lensed events without actually performing parameter estimation for a large number of signals, we need joint distributions $P(\rho_1, \rho_2, \dt \mid \HL)$ of time delays and S/N under each observing scenario. We follow the methodology from \cite{haris2018identifying, barsode2025fast}. We assume BBHs distributed according to the population from \cite{abbott2023population}, with two assumptions of merger rate distributions, namely \cite{madau2014cosmic} and \cite{dominik2013double}. Lens properties are taken from the SDSS catalog of galaxies \citep{collett2015population}, adopting the Singular Isothermal Ellipsoid (SIE) \citep{kormann1994isothermal, fukugita1991gravitational} profile.

We adopt data products from Maity et al. (in prep) regarding selection effects. These are simulated under various observing scenarios as described in Table~\ref{tab:scenario_definitions}. We assume each scenario runs for 3 years, and estimate the total number of detections $\ngw$ and the lensing fraction $u$ accordingly. We consider signals detectable if their optimal S/N exceeds 9, and require both images to occur within the 3-year observing period, where the arrival time of the first image is drawn uniformly.

The uncertainties in merger rates and lens models can substantially affect our conclusions on the detectability of lensed events. We analyze 18 combinations in total. For each of the two redshift models (\cite{madau2014cosmic, dominik2013double}), we vary the merger rate across the lower, median, and higher bounds inferred from \cite{abbott2023population}. For each of these, the lensing fraction is set to either 0.1\%, 1\%, or the value computed from our simulations. We summarize the range of outcomes across the 18 combinations in the following three categories:
\begin{itemize}
\item \emph{Realistic:} Local merger rate is taken as the median rate from \cite{abbott2023population} and is extrapolated using \cite{madau2014cosmic} redshift distribution. Lensing fractions are inferred using our simulation.
\item \emph{Pessimistic:} These correspond to the lowest efficiencies and numbers of confident identifications among all 18 combinations.
\item \emph{Optimistic:} These correspond to the highest efficiencies and numbers of confident identifications among all 18 combinations.
\end{itemize}

\begin{figure*}[t]
\centering
\includegraphics[width=\columnwidth]{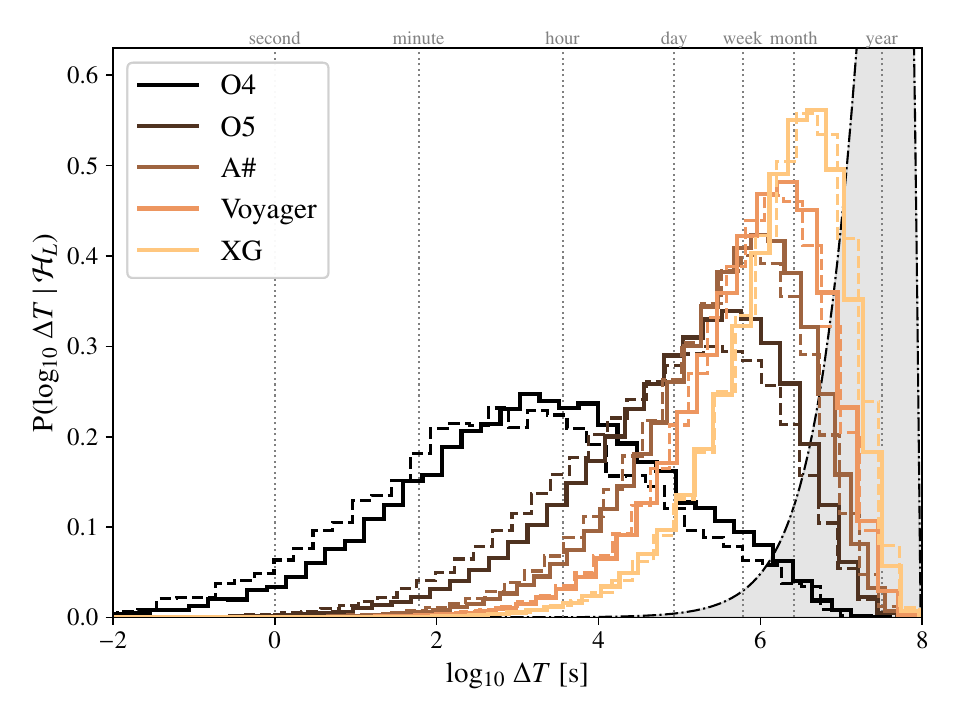}
\includegraphics[width=\columnwidth]{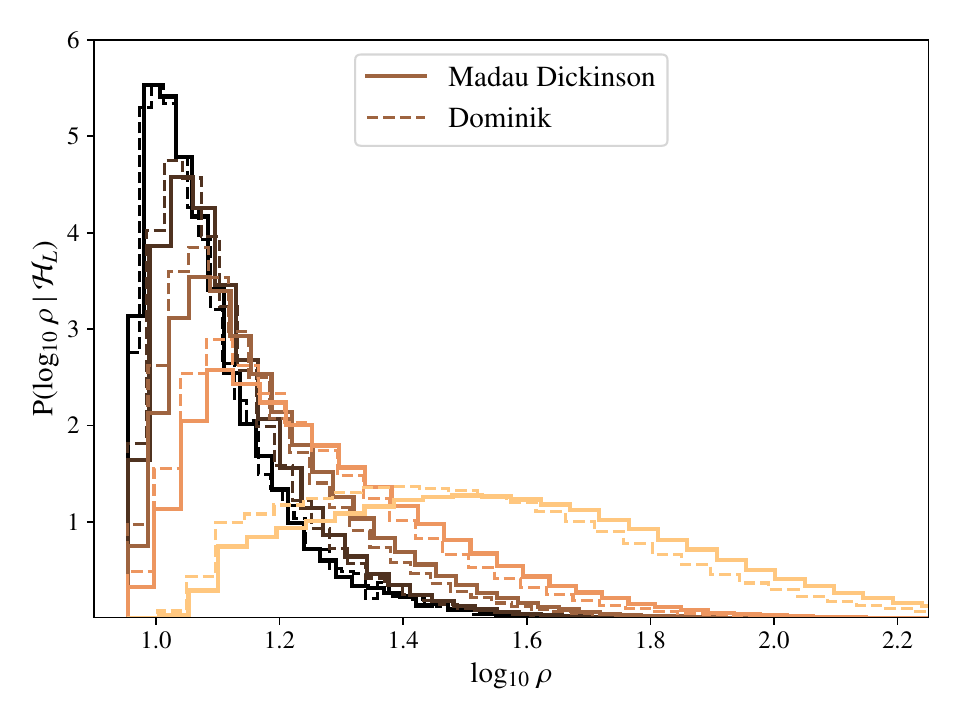}
\includegraphics[width=\columnwidth]{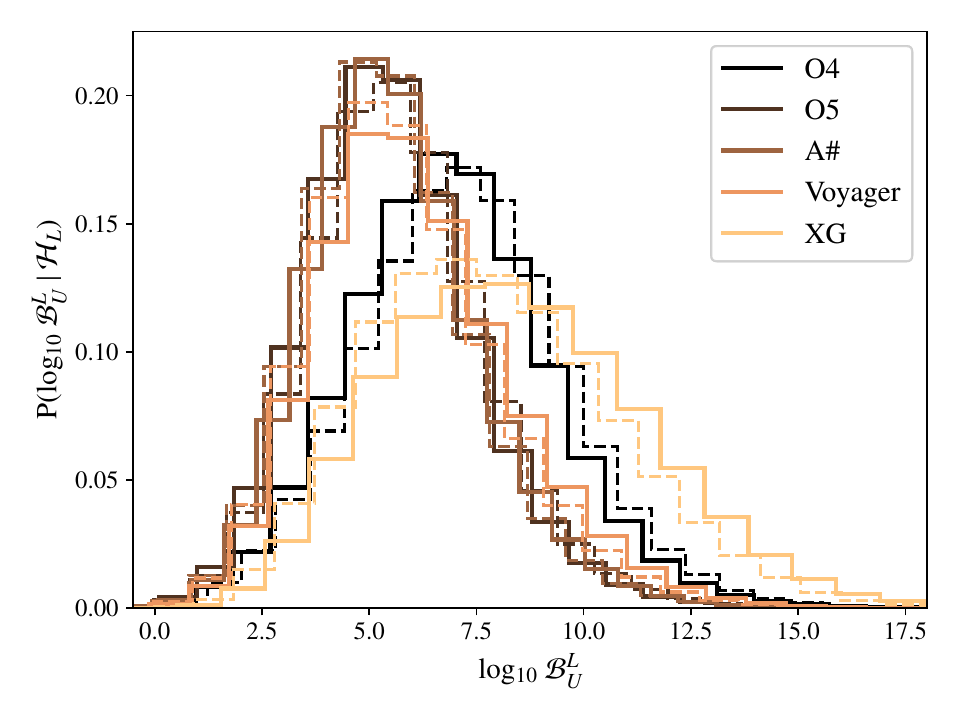}
\includegraphics[width=\columnwidth]{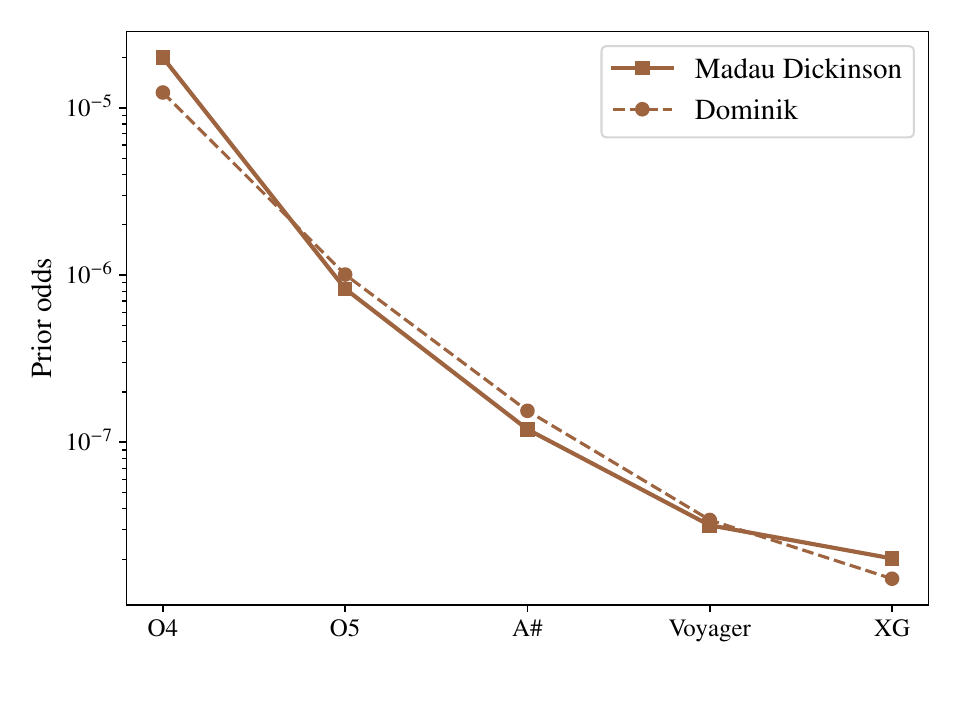}
\caption{\emph{Upper left}: The distribution of time delays of lensed images detectable under various observing scenarios. Results are shown for \cite{madau2014cosmic}'s and \cite{dominik2013double}'s redshift distributions for BBHs. The time delay distribution for Poisson distributed unlensed pairs is also shown in gray. \emph{Upper right}: The distribution of effective S/N $\rho$ for detectable lensed images. \emph{Lower left}: The distribution of $\blu$'s of lensed pairs. \emph{Lower right}: The prior odds ($= 2u/\ngw$) of encountering lensed pairs.}
\label{fig:P_dt_SNR0_BLU_SIE_KNM_scenarios}
\end{figure*}

The distributions of realistic time delays and S/N, $P(\dt \mid \HL)$, $P(\rho \mid \HL)$, are shown in the upper panels of Figure~\ref{fig:P_dt_SNR0_BLU_SIE_KNM_scenarios}. Intrinsically, time delays tend to be large \citep{jana2024strong}. However, magnifications are typically anticorrelated with time delays \citep{meneghetti2021introduction}, resulting in a selection bias towards highly magnified, lower time delay pairs. As these selection effects become less important due to increasing S/N, the time delay distribution tends to shift towards higher values, approaching its intrinsic distribution \citep{more2022improved}.

In the O4 scenario, we perform Bayesian parameter estimation runs on all detected events using the IMRPhenomXAS~\citep{pratten2020setting} waveform approximant, employing the \textsc{cogwheel} package~\citep{islam2022factorized, roulet2022removing}. The $\blu$'s are calculated using PO2.0~\citep{barsode2025fast}. In order to make the $B-B$ plot discussed in appendix~\ref{sec:sim_strong_lensing}, we also simulate an astrophysical population of unlensed BBHs detectable in O4, using the same prescription described above, sans the lensing part.

To compute the distribution of Bayes factors expected in \emph{future} observing runs, we follow the following steps:
\begin{enumerate}
\item Use the distribution of the effective S/N and time delays of lensed pairs, $P(\rho, \dt \mid \HL)$, to compute the the expected distribution of Bayes factor, $P(\blu \mid \HL)$, in a given future observing run, making use of the method described in Appendix~\ref{sec:sim_strong_lensing}. These are shown in the lower left panel of Figure~\ref{fig:P_dt_SNR0_BLU_SIE_KNM_scenarios}.
\item Use the relation given in Equation~\eqref{eq:blu_distribution_relationship_inverted} to compute the expected distribution of $\blu$ from unlensed events, i.e., $P(\blu \mid \HU)$.
\item Use Equation~\eqref{eq:FPPcat} to compute the threshold $\bt$ corresponding to an acceptable value of $\FPPcat$. Alternatively, use Equation~\eqref{eq:purity} to compute the threshold $\bt$ corresponding to an acceptable value of $\mathrm{Pty}$.
\item Use Equation~\eqref{eq:efficiency} to compute the lensing identification efficiency corresponding to the threshold $\bt$. Compute the number of detectable lensed events using Equation~\eqref{eq:Nlens_det}.
\end{enumerate}

The resulting values of lensing identification efficiency and expected number of lensed events are reported in Figures \ref{fig:future_efficiency_ROC} and \ref{fig:future_efficiency_numbers}.

\section{$\olu$ thresholds for different values of $\FPPcat$ and purity}
\label{sec:OLU_mapping}
\begin{figure}[t]
\centering
\includegraphics[width=\columnwidth]{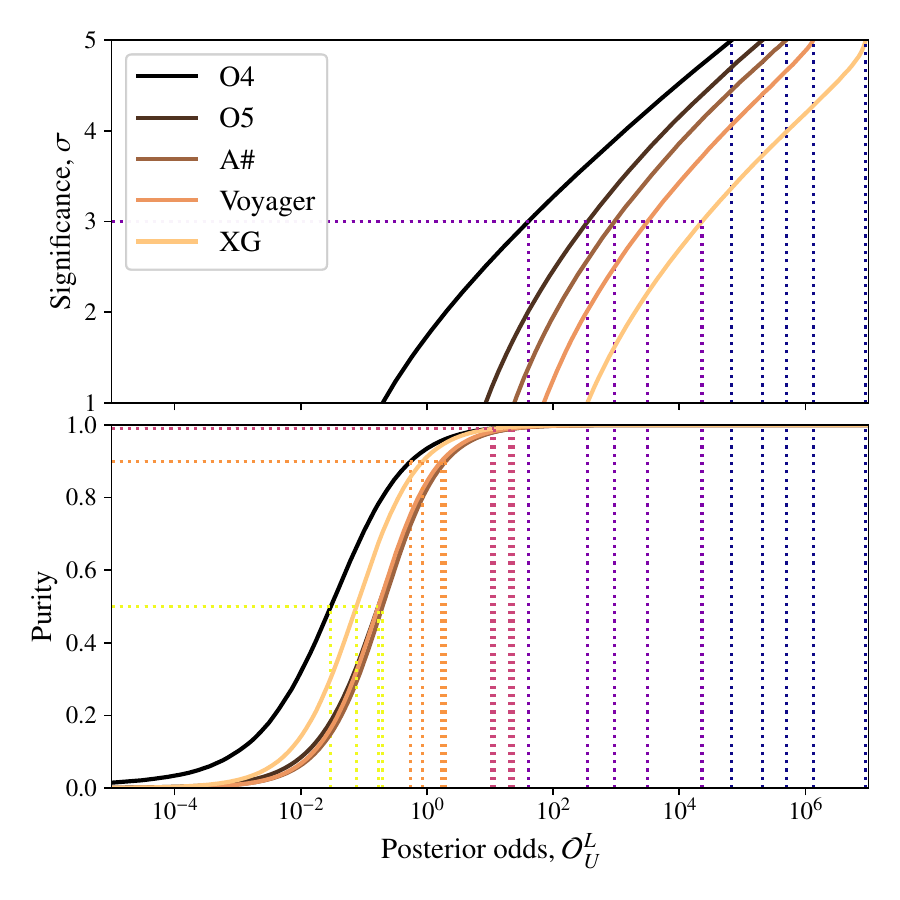}
\caption{The purity and frequentist $\FPPcat$ (in terms of Gaussian $\sigma$) plotted against the Bayesian posterior odds in various observing scenarios. These assume the \cite{madau2014cosmic} merger rate; the results with \cite{dominik2013double} are similar.}
\label{fig:posterior_odds_thresholds}
\end{figure}

The search for strongly lensed GWs is a classification problem between lensed and unlensed pairs. In the Bayesian approach, this is based on the largeness of the posterior odds $\olu$, given by the product of prior odds and the Bayes factor $\blu$. On the other hand, when the frequentist statistic is chosen to be $\olu$ or $\blu$\footnote{Note that the frequentist $\FPPpair$ is a cumulative distribution function that is unchanged if the statistic is scaled by a constant, such as $\olu \propto \blu$, so long that the thresholds are also scaled appropriately.}, the classification criteria technically remain the same. Thus the lensed catalogs that would be generated by the two approaches would be identical for the same thresholds $\ot=\Pi^L_U ~ \bt$.

What may differ in the two approaches is the level of confidence assigned to each pair in the lensed catalog. This confidence is quantified by the probability that labelling a pair as ``lensed'' might have been a mistake. In Bayesian, it is the probability $1/(1+O)$ that a pair with posterior odds $O$ may be unlensed, while in frequentist, it is the probability $\FPPcat(\ot=O)$ that an unlensed pair may have odds greater than $O$. Clearly, these two measures of confidence are not equal to each other. This is not surprising as such, while both of these provide a probability of making a mistake, the exact questions they are answering are completely different.

In Figure~\ref{fig:posterior_odds_thresholds}, we show the mapping between catalog purity and frequentist $\FPPcat$ (expressed in terms of Gaussian $\sigma$) against posterior odds $\olu$ for different observing scenarios. The threshold $\olu$ may vary by over 2 orders of magnitude at a given $\sigma$. Note that this is true even under controlled conditions where the correspondence from Equation~\eqref{eq:blu_distribution_relationship_inverted} holds. This is because $\FPPcat$ and $\sigma$ depend not only on the $\olu$ itself, but also on its probability density, which may differ between different scenarios. The variation between threshold $\olu$ is relatively lower at constant purity, and, if a background is unavailable, one may roughly take posterior odds of 1 as a threshold corresponding to a catalog purity of 90\% with the caveat that $\olu$ may be biased due to various model uncertainties.

\bibliography{future_ROC}

\end{document}